\documentclass[adp,fleqn]{w-art}
\usepackage{times}
\usepackage{w-thm}
\usepackage[]{graphicx}
\chardef\bslash=`\\ 

\begin{document}
\Volume{13}
\Year{2004}
\pagespan{1}{}


\title[Effective quantum theories and path integrals]{Effective
       descriptions of complex quantum systems: \\ path integrals
       and operator ordering problems}

\author[U. Eckern]{Ulrich Eckern\footnote{Corresponding author\quad
     E-mail: {\sf ulrich.eckern@physik.uni-augsburg.de}}}

\address[]{Institute of Physics, University of Augsburg, 86135 Augsburg,
           Germany}
\author[M. J. Gruber]{Michael J. Gruber}
\author[P. Schwab]{Peter Schwab}

\begin{abstract}
  We study certain aspects of the effective, occasionally called
  collective, description of complex
  quantum systems within the framework of the path integral formalism,
  in which the environment is integrated out. Generalising the
  standard Feynman-Vernon Caldeira-Leggett model to 
  include a non-linear
  coupling between ``particle'' and environment, and considering a
  particular spectral density of the coupling, a coordinate-dependent
  mass (or velocity-dependent potential) is obtained. The related
  effective quantum theory, which depends on the proper discretisation
  of the path integral, is derived and discussed. As a result, we find
  that in general a simple effective low-energy Hamiltonian, in
  which only the coordinate-dependent mass enters, cannot be formulated.
  The quantum theory of
  weakly coupled superconductors and the quantum dynamics of vortices 
  in Josephson junction arrays are physical examples where these
  considerations, in principle, are of relevance.
\end{abstract}
\maketitle                   

\section{Introduction}
\label{intro}

Since the seminal contributions by Richard Feynman in 1948, it is clear
that the description of quantum
and quantum statistical systems can be based either on wave-functions and
the Schr\"odinger equation, or on the path integral prescription for
computing the propagator and/or the partition function. Both descriptions
are equivalent: Given a certain Schr\"odinger equation, the corresponding
path integral can be formulated, and, alternatively, a path integral 
expression can be used to define a quantum theory. In particular, the
latter route appears to have advantages in complex many-body systems
and for field theories. These topics are, of course, well documented in
several textbooks, for example, by Feynman and Hibbs \cite{feynman65},
Schulman \cite{schulman81}, Sakita \cite{sakita85}, and Kleinert
\cite{kleinert95}. In order to set the stage,
consider e.\,g.\ the partition function of a simple quantum system,
\begin{equation}
\label{eq1}
Z = \int {\cal D}x \exp (- S_E / \hbar) \; ,
\end{equation}
where $S_E$ is the Euclidean action,
\begin{equation}
\label{eq2}
S_E [x(\tau)]=\int_0^\beta d\tau \Big[ \frac{m}{2}{\dot x}^2+V(x)\Big] \; .
\end{equation}
As usual, $m$ is the mass of the particle, $V(x)$ the potential, and
${\dot x} \equiv \partial x / \partial\tau$. The symbol ${\cal D}x$ is
a short-hand notation for a multi-dimensional integral, defined through
a discretisation of the interval $0 \dots \beta$, by putting
$\tau_j = j \epsilon$, $j = 0, 1, 2, \dots N$ ($\beta = N\epsilon$), and
$x(\tau_j) = x_j$, in the limit $N\to\infty$, $\epsilon\to 0$ so that
$\beta = \hbar /k_B T$ (where $T$ is the temperature) remains finite. With
an appropriate deformation of the time-integration contour, and adding 
appropriate source fields, Eq.\ (\ref{eq1}) can be generalised to a
generating functional for real-time correlation functions, e.\,g., in
analogy to the diagrammatic technique due to Keldysh \cite{keldysh65}.
Note that, when evaluating the multi-dimensional integral in Eq.\ (\ref{eq1}),
the correct integration measure is essential -- and that appropriate paths
have to be taken into account: for the propagator, $x_0$ and $x_N$ are to
be kept fixed, while for the partition function, $x_0 = x_N$ with an
additional integration with respect to this variable (since 
$Z$ is a trace).

In a more general situation, say, in which the potential is velocity
dependent (as for a charged particle)
it is essential to choose
the correct discrete representation, which is known as the mid-point
prescription, i.\,e.\ 
\begin{equation}
\label{eq3}
\int d\tau {\dot x} a (x) \to \epsilon\sum_j \frac{(x_j - x_{j-1})}{\epsilon} 
     a( \frac{x_j + x_{j-1}}{2} ) \; .
\end{equation}     
This prescription ensures that the corresponding Hamiltonian 
operator contains the symmetric
form ${\hat p} a( {\hat x} ) + a( {\hat x}) {\hat p}$, and hence its
hermiticity. (The physically relevant case for
this example is, of course, in three dimensions.) A closely related
problem appears in the theory of Brownian motion, in particular, for
the Ornstein-Uhlenbeck process, defined by the Langevin equation
\begin{equation}
\label{eq4}
\eta {\dot x} = -kx + \xi (t) \; ,
\end{equation}
where $\xi (t)$ represents Gaussian white noise (with zero average). 
Considering the generalisation where $-kx$ is replaced by some function
$f(x)$, and starting from the Gaussian probability density for the noise,
it is straightforward to derive the probability density for the random 
process $x(t)$, $W(\{x\})$, with the help of Eq.\ (\ref{eq4}):
\begin{equation}
\label{eq5}
W(\{x\}){\cal D}x = W(\{\xi\}){\cal D}\xi \; .
\end{equation}
The result is unique, as is the corresponding Fokker-Planck equation; the
intermediate steps, however, depend on the discretisation procedure, the
two cases discussed in this context being connected with the names
Ito (forward rule) and 
Stratonovich (mid-point rule) \cite{kubo91}. In the former, the
Jacobian of the transformation is a mere constant, but the integral of
${\dot x}f(x)$ has a non-trivial contribution; in the latter, the
integral of ${\dot x}f(x)$ depends only on the end-points -- but care is
needed to evaluate the Jacobian. Physically, this difficulty is related
to the irregularity of the Brownian motion, and hence can be cured, for
example, by introducing a finite mass term $m{\ddot x}$ into Eq.\
(\ref{eq4}). A concise discussion of the question of operator ordering
and functional formulations in both quantum and stochastic dynamics
can be found in \cite{leschke77}; the arguments given above, to the
best of our knowledge, were first formulated by Schmid \cite{schmid82}
(see also \cite{eckern90}). 

Until now, we implicitly assumed that the Hamilton operator of the
system in question is given -- and we will make the same assumption in
the following sections, as we have in mind applications 
to condensed matter physics.
We would mention, however, that a more general question is
(still) studied intensively in mathematical physics: Given a classical
Hamiltonian function, what is the ``correct'' quantum theory? For an
introduction to related topics, see, for example, \cite{waldmann02}; in
this connection, the concept of Weyl ordering (\cite{sakita85},
chap. VI, and \cite{waldmann02}; see also below) is useful: as is well
known, a Weyl ordered Hamiltonian corresponds to a path integral
defined by the mid-point rule. In complex
many-body systems, a related question shows up in a natural way when
formulating an effective quantum theory by integrating out a
subset of the variables -- and this is the focus of the present paper.

In the next section, Sect. \ref{model}, we briefly recapitulate the
essence of the Feynman-Vernon \cite{feynman63} Caldeira-Leggett
\cite{caldeira81} model, which is based on the assumption that, in
a complex many-body system, a suitable coordinate (``particle'') can be
identified; and that this particle is weakly coupled to the remaining
degrees of freedom (``bath''). The choice of the coupling and the bath,
the latter typically assumed to be a set of harmonic oscillators, depends on
the problem under consideration: in particular, the choice which corresponds
to an effective dissipation has attracted considerable attention
(see \cite{leggett87} and \cite{schoen90} for reviews). The following Sect. 
\ref{super} is devoted to weakly coupled superconductors where, as
shown from microscopic theory \cite{ambegaokar82,larkin83, ambegaokar84},
the coupling between particle (i.\,e.\ the order parameter phase) and
environment (i.\,e.\ the electronic degrees of freedom)
can be considered as non-linear, 
reflecting the $2\pi$-periodicity of the phase variable \cite{ambegaokar87}.
As a result, a phase-dependent mass (i.\,e.\ capacitance in this case)
is found in the effective action \cite{larkin83, ambegaokar84}; see also
\cite{shimshoni89,eckern88}. In the central Sect. \ref{eff}, we investigate
in detail one of the simplest models corresponding -- in the classical
limit and for low frequencies -- to a coordinate-dependent 
mass in the effective action.
We briefly review some classical aspects, and discuss the relation
between operator ordering and path integral formulation. Using the
(discrete) path integral formulation as well as perturbation theory and
a variational approach, we then discuss the question whether,
in general, an effective low-energy quantum description exists which 
can be expressed in terms of the coordinate-dependent mass.
The conclusions are summarised
in Sect. \ref{con}.

\section{The paradigmatic model}
\label{model}
The Feynman-Vernon Caldeira-Leggett \cite{feynman63,caldeira81} (FV-CL)
model is defined through the following Hamilton operator:
\begin{equation}
\label{eq6}
{\hat H} = \frac{{\hat p}^2}{2m} + V({\hat x}) +
         \sum_\mu \Big[ \frac{{\hat p}_\mu^2}{2} +
	 \frac{\omega_\mu^2}{2} \Big( {\hat x}_\mu - 
	 \frac{C_\mu}{\omega_\mu^2} \hat x \Big)^2 \Big]
\end{equation}
where $\hat p$, $\hat x$ and ${\hat p}_\mu$, ${\hat x}_\mu$ denote
the momentum and position operators of the particle and the environment,
respectively, the latter being labeled by $\mu$. The masses of the
oscillators, without loss of generality, have been put equal to unity,
and the $\{C_\mu\}$ are the particle-environment coupling constants.
We have chosen the form of the coupling above so that certain counter
terms are avoided. For the next step, it is important that
the coupling, $\hat x\sum_\mu C_\mu {\hat x}_\mu$, is linear in
the bath variables $\{{\hat x}_\mu\}$, in order to allow them to be integrated
out easily. (We could, of course, have chosen a coupling to the momenta
$\{{\hat p}_\mu\}$ as well.) Performing the integration, one finds
an effective action which contains the particle variable only; the
properties of the bath and the particle-bath coupling are absorbed in
a certain correlation function. The result is 
\begin{equation}
\label{eq7}
S_E^{{\rm (eff)}} = S_E^{(0)} + S_E^{(1)} \; ,
\end{equation}
where $S_E^{(0)}$ is given
by Eq.\ (\ref{eq2}), and
\begin{equation}
\label{eq8}
S_E^{(1)} = \frac{1}{2} \int_0^\beta \! d\tau \int_0^\beta \! d\tau^\prime
     \alpha (\tau - \tau^\prime) \Big[ x(\tau) - x(\tau^\prime) \Big]^2 \; .
\end{equation}
Here the function $\alpha (\tau)$, which is an even function of time, 
is given by
\begin{equation}
\label{eq9}
\alpha (\tau) = (2\beta)^{-1} \sum_{\omega} {\rm e}^{-i\omega\tau}
              \sum_\mu \frac{C_\mu^2}{\omega^2 + \omega_\mu^2}
\end{equation}
and has the following alternative representation:
\begin{equation}
\label{eq10}
\alpha (\tau) = \int_{-\infty}^{+\infty} \frac{d\nu}{2\pi}
      J(\nu) [-b(-\nu)] {\rm e}^{-\nu |\tau|}
\end{equation}
where $J(\nu)$ is given by
\begin{equation}
\label{eq11}
J(\nu) = \frac{\pi}{2} \sum_\mu \frac{C_\mu^2}{\omega_\mu}
     \Big[ \delta (\nu - \omega_\mu)-\delta (\nu + \omega_\mu)\Big] \; .
\end{equation}
Above, $\omega = 2\pi n/\beta$ are the 
Matsubara frequencies, and $b(\cdot)$ is
the Bose function. The dissipative case is realised by choosing an infinite
set of bath oscillators with a dense frequency distribution, so that
$J(\nu) = \eta\nu$ for small $\nu$, which corresponds to
$\alpha (\omega) = \alpha (0) -(\eta/2) |\omega|$ and, in
the zero-temperature limit ($\beta\to\infty$), to $\alpha (\tau) = 
(\eta/2\pi) \tau^{-2}$. The parameter $\eta$ is the viscosity,
entering the classical (real-time) equation of motion in the form
$m{\ddot x} + \eta{\dot x} = \dots$; for further details, see, for example,
\cite{caldeira81} and \cite{caldeira80}. Note the close correspondence
of $J(\nu)$ to the function $\alpha^2 F(\nu)$, well-known from the
theory of superconductivity \cite{scalapino69}.

In contrast, consider a situation where there is a gap in the spectrum
of the bath oscillators, so that $J(\nu) = 0$ for $|\nu| < \Delta$.
Then $\alpha (\omega) = \alpha (0) -(m/2)\lambda \omega^2$ for small
frequencies, thereby defining the parameter $\lambda$ ($>0$).
This implies a mass
enhancement according to $m \to m (1+\lambda)$, well-known from the
polaron problem; compare \cite{feynman65},
Chap. 11-4, or \cite{schulman81}, Chap. 21. Explicit results are easily
obtained for the simplest case of a {\em single} bath oscillator.

Generalisations can be obtained by considering a non-linear coupling 
between particle and environment, i.\,e.\ by replacing ${\hat x}$ by
a non-linear function $g({\hat x})$ in the last term of Eq.\ (\ref{eq6}). 
Furthermore, several environments
can be introduced, for convenience chosen to be independent of each other,
so that 
\begin{equation}
\label{eq12}
\{ {\hat p}_\mu, {\hat x}_\mu\} \to 
\{{\hat p}_\mu^{(m)}, {\hat x}_\mu^{(m)} \} \; , \;\;
g({\hat x}) \to g^{(m)} ({\hat x}) \; .
\end{equation}
As a result, Eq.\ (\ref{eq8}) generalises to the following expression:
\begin{equation}
\label{eq13}
S_E^{(1)} = \frac{1}{2} \sum_m\int_0^\beta \! d\tau\int_0^\beta \! d\tau^\prime
     \alpha^{(m)} (\tau - \tau^\prime) 
     \Big[ g^{(m)}(x(\tau)) - g^{(m)}(x(\tau^\prime)) \Big]^2 \; .
\end{equation}
It is clear that in general, an equivalent Hamiltonian description does
not exist; 
we return to this question in Sect. \ref{eff}.

\section{Weakly coupled superconductors}
\label{super}
In this section, we briefly review the steps which lead to an effective
quantum description of weakly coupled superconductors, i.\,e.\ Josephson
junctions \cite{ambegaokar82,larkin83,ambegaokar84}, and recall how the
effective action also can be derived within the generalisation of the
FV-CL model as given by Eq.\ (\ref{eq13}). The starting point is the BCS
model of superconductivity, i.\,e.\ electrons which weakly attract each 
other. Two of these BCS superconductors are weakly coupled by the
tunneling Hamiltonian \cite{josephson62}, 
which allows for electron transfer from one
superconductor to the other. In the first step, within the path integral
formulation, complex order parameter fields are introduced to decouple the
attractive interaction. Similarly, the Coulomb interaction 
between charges near the
junction surfaces, which becomes an effective capacitive
interaction, is described by a voltage field. In an intermediate step,
we obtain a form which is bilinear in the four-dimensional (left/right,
spin-up/spin-down) space of the fermionic (Grassmann) variables.
In the second step, the fermionic fields are integrated out. In addition,
it is advantageous to perform a gauge transformation which makes the
particle-hole off-diagonal elements real and reveals 
the role of the order parameter phases. In a third step, the action is
expanded (up to second order) in the tunnel matrix elements. Based on
a detailed analysis of the different contributions one can
show that the effective low-frequency action, which is a functional
of the phase difference $\phi$ across the junction, is given by
\begin{equation}
\label{eq14}
S_E^{\rm (JJ)} = S_E^{\rm (C)} + S_E^{\rm (T)} 
\end{equation}
where
\begin{equation}
\label{eq15}
S_E^{\rm (C)} = \int_0^\beta d\tau \Big[ \frac{\hbar^2 C}{8e^2}
              {\dot\phi}^2 - \frac{\hbar I}{2e} \phi \Big]
\end{equation}
contains the capacitive energy (capacitance: $C$) as well as the
contribution from the external current ($I$), and
\begin{equation}
\label{eq16}
S_E^{\rm (T)} = \hbar \int_0^\beta \! d\tau\int_0^\beta \! d\tau^\prime
    \Big[ - \alpha (\tau - \tau^\prime)
    \cos \frac{\phi(\tau) - \phi(\tau^\prime)}{2} 
    + \beta (\tau - \tau^\prime)
    \cos \frac{\phi(\tau) + \phi(\tau^\prime)}{2} \Big] \; .
\end{equation}
Here we use the same notation as in \cite{ambegaokar82}. The kernels
$\alpha(.)$ and $\beta(.)$ are related to the diagonal and off-diagonal
(equilibrium) Green's functions of the superconductors, and hence can
be expressed through the functions $I_n (\nu)$ and $I_c (\nu)$
which characterise the quasiparticle current and the supercurrent across
the junction \cite{werthamer66} (see also \cite{ambegaokar87}).
Considering zero temperature and slow variations (compared to the
inverse gap frequency) in time, we may use the expansions 
$\alpha (\omega) = \alpha_0 - \alpha_2 \omega^2 /2$ and
$\beta (\omega) = - \beta_0 + \beta_2 \omega^2 /2$ (where 
$\alpha_2$, $\beta_0$, $\beta_2$ are all positive; $\alpha_0$ is
irrelevant) to show that the $\beta_0$-term leads to the usual
cosine potential, $-E_J \cos\phi$, where $E_J = \pi\hbar\Delta/4e^2R_N$ 
is the Josephson coupling
energy; and that the $\alpha_2$- and $\beta_2$-contributions
change the capacitance (i.\,e.\ the ``mass'') according to
\begin{equation}
\label{eq17}
C \to C(\phi) = C + C_{\rm qp} \Big( 1 - \frac{1}{3} \cos\phi \Big)
\end{equation}
where $C_{\rm qp} = 3\pi\hbar /(32\Delta R_N)$; $\Delta$ is the
superconducting gap, assuming the two superconductors to be equal,
and $R_N$ the normal-state resistance of the junction.

The above model, often -- for simplicity -- considered in the limit
$C_{\rm qp} \ll C$, is also a good starting point for discussing the quantum
properties of an array of Josephson junctions, for example the quantum
dynamics of a vortex \cite{eckern89}: in this case, a position-dependent
mass arises due to the lattice structure of the underlying array of
superconducting grains.\footnote{In the
  absence of additional shunt resistors,
  and for zero external current, it is important that the phase variable
  is defined on the interval $0\dots 2\pi$, with $0$ and $2\pi$ to be
  identified, thus requiring a different interpretation of the path
  integral; see e.\,g.\ \cite{schulman81}, Chap. 23.}

Considering again Eq.\ (\ref{eq13}), it is apparent that the effective
action of a Josephson junction, Eq.\ (\ref{eq16}), can be ``derived''
from the generalisation of the FV-CL model by an appropriate choice of
$g^{(m)}(\cdot)$ and $\alpha^{(m)}(\cdot)$, in particular, 
using $g^{(1)}(\phi) = \sin (\phi/2)$, and
$g^{(2)}(\phi) = \cos (\phi/2)$; see \cite{ambegaokar87} for
details. 

\section{Coordinate-dependent mass and effective Hamiltonian}
\label{eff}
In Sect. \ref{model} we have shown that a coordinate-dependent mass 
can be considered as originating from
a quite simple model, namely a particle coupled non-linearly
to a harmonic oscillator of finite frequency.\footnote{Some aspects of
  this model were discussed a while ago \cite{leschke84}. In
  particular, it was pointed out that
  it is legitimate to define an effective Hamiltonian by tracing
  over some of the system's variables.}
Hence we will study in this
section the following Hamilton operator in more detail:
\begin{equation}
\label{eq19}
{H} = \frac{{\hat p}^2}{2m} + V({\hat x}) +
         \frac{1}{2} \Big[{\hat p}_\mu^2 + \omega_\mu^2 {\hat x}_\mu^2 \Big]
	 + g({\hat x}) {\hat x}_\mu + \frac{g^2 ({\hat x})}{2\omega_\mu^2}
	 \; .
\end{equation}
The subscript ``$\mu$'' is kept for easy distinction. As an alternative,
we may perform a unitary transformation \cite{lang63}: 
Defining $U = \exp [i g({\hat x})
{\hat p}_\mu /\hbar\omega_\mu^2 ]$, we obtain
\begin{equation}
\label{eq19a}
{\tilde H} \equiv U^+ H U = 
         \frac{1}{2m} \Big( {\hat p} + a({\hat x},{\hat p}_\mu) \Big)^2
         + V({\hat x}) +
         \frac{1}{2} \Big[{\hat p}_\mu^2 + \omega_\mu^2 {\hat x}_\mu^2 \Big]
\end{equation}
where $a({\hat x},{\hat p}_\mu)
 = g^\prime ({\hat x}){\hat p}_\mu /\omega_\mu^2$. In ${\tilde H}$, the
coupling is to the momentum of the oscillator, and involves directly the
derivative of the coupling function, $g^\prime ({\hat x})$ (which turns
out to be useful). We also note that in the derivation of the effective
action on the basis of the transformed Hamiltonian, it is convenient to
consider the phase-space representation of the path integral, and to
integrate first with respect to the particle momenta.

\subsection{The polaronic toy model: classical aspects}

In order to derive the classical low-frequency effective action of the
model defined by Eq.\ (\ref{eq19}) or Eq.\ (\ref{eq19a}), we 
concentrate on the zero-temperature limit where
$\alpha(\tau) = (4\omega_\mu)^{-1}
\exp (-\omega_\mu |\tau|)$, and consider slow variations in time
(compared to $\omega_\mu^{-1}$) so that we may expand the effective
action as follows:
\begin{equation}
\label{eq20}
S_E^{(1)} = \frac{1}{2} \int \! d\tau \int \! d\tau^\prime
     \alpha (\tau - \tau^\prime)
          \Big[ g(x(\tau)) - g(x(\tau^\prime)) \Big]^2 
           \approx \frac{\alpha_2}{2} \int \! d\tau \,
           [g_1 (x)]^2 \, {\dot x}^2
\end{equation}
where $g_1 (x) = dg(x)/dx$; the constant $\alpha_2 = - 
[\alpha^{\prime\prime} (\omega)]_{\omega\to 0}$
is given by $\omega_\mu^{-4}$. For $g_1 (x) \equiv 1$, the mass enhancement
is simply given by the factor $1 + \alpha_2 /m$, as discussed in Sect.\
\ref{model}. The last expression in (\ref{eq20}) follows by noting that the
integrand in the effective action is strongly peaked for 
$\tau\approx\tau^\prime$, allowing for the approximations
\begin{equation}
\label{eq21}
x(\tau) \approx x(\frac{\tau +\tau^\prime}{2}) + \frac{\Delta\tau}{2}
 {\dot x} (\frac{\tau +\tau^\prime}{2}) \; , \;
x(\tau^\prime) \approx x(\frac{\tau +\tau^\prime}{2}) - \frac{\Delta\tau}{2}
 {\dot x} (\frac{\tau +\tau^\prime}{2}) \; , 
\end{equation} 
where $\Delta\tau = \tau -\tau^\prime$ denotes the difference time,
and also expanding $g(x(\tau))$ and $g(x(\tau^\prime))$.

Clearly, the effective action of this ``polaronic toy model'' corresponds
to the following classical Lagrangian and Hamiltonian:
\begin{equation}
\label{eq22}
{\cal L} = \frac{1}{2} m(x) {\dot x}^2 - V(x) \;\; \leftrightarrow \;\;
{\cal H} = \frac{p^2}{2m(x)} + V(x) \; .
\end{equation}
with $m(x) = m + \alpha_2 [g_1 (x)]^2 )$ denoting the 
coordinate-dependent mass, and ${\dot x} = dx/dt$.
The canonical momentum is $p=m(x){\dot x}$, and
the classical equation of motion reads
\begin{equation}
\label{eq23}
m(x){\ddot x} + \frac{1}{2} m^\prime (x) {\dot x}^2 = - V^\prime (x) \; .
\end{equation}
Of course, the energy, $E = m(x){\dot x}^2 /2 + V(x)$, is conserved. In
the present -- one-dimensional -- case (see also \cite{sakita85} 
and \cite{kleinert95}, as well as \cite{kleinert02} and \cite{koc03}
for related discussions of the general case) it is obvious that the
above expressions can be ``simplified'' by the coordinate transformation
${z}={z}(x)$, defined by 
$m(x){\dot x}^2=\bar{m}{\dot z}^2$
\cite{eckern88,eckern89}. The equation of motion then
reads $\bar{m} \ddot{z} =-d U(z)/dz$, 
where $U ({z}) \equiv V(x({z}))$. Clearly, we have to assume $m(x)>0$ in
order to assure stability; and we will assume below that $m(x)$ is
differentiable. 

The example discussed in Sect.\ \ref{super}, i.\,e.\
weakly coupled superconductors, corresponds to $m(x) = m_0 + m_1 \cos x$
with $m_0 = (\hbar^2/4e^2)(C + C_{\rm qp})$, 
$m_1 = - (\hbar^2/4e^2) C_{\rm qp} /3$, and
the momentum to $\hbar /2e$ times the charge. In addition,
$V(x) = -E_J \cos\phi - (\hbar I/2e)\phi$; in case the Josephson contact
is embedded in a superconducting ring, instead of the term $\sim\phi$, a
quadratic contribution $(\phi - \phi_{\rm ext})^2 /2L$ 
arises \cite{tinkham96} where
$\phi_{\rm ext}$ and $L$ are proportional to the external magnetic flux
and the inductance of the ring, respectively.

\subsection{Coordinate-dependent mass: some remarks on quantisation}
The above considerations suggest a straightforward route to quantise a
system with coordinate-dependent mass \cite{sakita85,kleinert95,kleinert02},
namely to consider the $z$-version 
${H}_z = {\hat p}_z^2 /2\bar{m} + U({\hat z})$ 
with $[{\hat p}_z , {\hat z}] =\hbar/i$.
Transforming back to the $x$-representation, the change in the
integration measure (consider, e.\,g., a scalar product) has to be
taken into account, since $dx = [\bar{m} / m(x)]^{1/2} dz$. The result
is
\begin{equation}
\label{eq24}
{H}_{\rm pct} =
\frac{1}{2}\frac{1}{m^{1/4}}{\hat p}\frac{1}{m^{1/2}}
{\hat p}\frac{1}{m^{1/4}}
+ V({\hat x})
\end{equation}
where $[{\hat p} , {\hat x}] =\hbar/i$, and the subscript indicates
``point canocical transformation''. Of course, the point canonical
transformation also can be considered within the path integral
formulation \cite{sakita85}: Starting from the standard expression for
the short-time propagator 
$\langle z_j |\exp (-\epsilon{\hat H}_z /\hbar ) |z_{j-1} \rangle$,
namely
\begin{equation}
\label{eq25}
K_z (z_j,\epsilon |z_{j-1},0)=
  \Big( \frac{\bar{m}}{2\pi\hbar\epsilon} \Big)^{1/2}
  \exp \Big\{ -\frac{\epsilon}{\hbar} \Big[
  \frac{\bar{m}}{2\epsilon^2} (z_j - z_{j-1})^2 +U(z_{j-1})\Big] \Big\} \; ,
\end{equation}
the short-time propagator in $x$-representation is given by
\begin{equation}
\label{eq26}
K_{\rm pct}(x_j,\epsilon | x_{j-1},0) = 
\Big[ \frac{m(x_j)m(x_{j-1})}{\bar{m}^2} \Big]^{1/4}
K_z (z(x_j),\epsilon | z(x_{j-1}),0) \; .
\end{equation}
Using the coordinate representation, and taking the limit $\epsilon\to 0$,
it is straightforward to confirm the Hamiltonian (\ref{eq24}).

Alternatively, the concept of Weyl ordering can be introduced: It is
well-known that a Weyl ordered Hamilton operator corresponds to the
mid-point description of the path integral \cite{leaf68}. Considering
a polynomial expression ${\hat p}^n {\hat x}^m$, the Weyl ordered form
$({\hat p}^n {\hat x}^m)_W$ is obtained by summing over all possible
orders, and dividing by their total number. In particular, this implies
\begin{equation}
\label{eq27}
\Big( {\hat p} f({\hat x}) \Big)_{\rm W} = \frac{1}{2} \Big(
{\hat p} f({\hat x}) + f({\hat x}) {\hat p} \Big)
\end{equation}
and
\begin{equation}
\label{eq28}
\Big( {\hat p}^2 f({\hat x}) \Big)_{\rm W} = \frac{1}{4} \Big(
{\hat p}^2 f({\hat x}) + f({\hat x}) {\hat p}^2 \Big)
+ \frac{1}{2} \Big( {\hat p} f({\hat x}) {\hat p} \Big) \; ,
\end{equation}
provided a meaningful power series expansion of $f({\hat x})$ exists.
This leads to
\begin{equation}
\label{eq29}
{H}_{\rm W} = \frac{1}{8m}{\hat p}^2 
+ {\hat p} \frac{1}{4m} {\hat p} + {\hat p}^2 \frac{1}{8m} 
+ V({\hat x}) \; .
\end{equation}
The pct-Hamiltonian differs from the Weyl Hamiltonian, with the
result \cite{sakita85}
\begin{equation}
\label{eq30}
{H}_{\rm pct} - {\hat H}_{\rm W}
= \frac{\hbar^2}{32} \frac{[m^\prime({\hat x})]^2}{m({\hat x})^3} 
\equiv {\cal W}({\hat x}) \sim \hbar^2 \; .
\end{equation}
Note that the difference has a definite sign. It was suggested by
Ambegaokar \cite{ambegaokar84} that the Weyl ordered form (\ref{eq29}) 
is the correct one
for the quantum theory of a Josephson junction.

Defining the mid-point ${\bar x}_j = (x_j + x_{j-1})/2$, we conclude
that the short-time (small $\epsilon$) propagator corresponding to
${\hat H}_{\rm W}$ is given by
\begin{equation}
\label{eq31}
K_{\rm W}(x_j,\epsilon | x_{j-1},0) = \int\! \frac{dp}{2\pi\hbar} \,
{\rm e}^{ip(x_j - x_{j-1})/\hbar} \,
{\rm e}^{-\epsilon {\cal H}(p,{\bar x}_j)/\hbar}
\end{equation}
with ${\cal H}$ defined in Eq.\ (\ref{eq22}); clearly
\begin{equation}
\label{eq32}
K_{\rm pct}(x_j,\epsilon | x_{j-1},0) =
K_{\rm W}(x_j,\epsilon | x_{j-1},0) \cdot 
\exp[-\epsilon {\cal W}({\bar x}_j) /\hbar] \; .
\end{equation}
Note that ${\bar x}_j$ can be replaced by $x_j$ in the last exponential
factor, as well as in the potential energy. After integration with respect
to the momentum, a ``non-trivial'' prefactor $\sim [m({\bar x}_j)]^{1/2}$
arises; this prefactor ensures that the propagators reduce to a
$\delta$-function in the limit $\epsilon\to 0$.

\subsection{The polaronic toy model: discrete path integrals}
In order to determine the correct quantum theory of the simplest model
of a particle coupled to an environment, we study the toy model
Eq.\ (\ref{eq19}) in more detail. The notation has been introduced in 
Sect.\ \ref{intro}, and everything is straightforward:
We retain the discrete version of the
path integral in the coordinate representation, 
and then ``integrate out'' the oscillator degrees
of freedom, with the following result:
\begin{equation}
\label{eq33}
Z / Z_{\mu} = \lim_{N\to\infty} 
\Big( \frac{m}{2\pi\hbar\epsilon} \Big)^{N/2}
\int\! dx_1 \dots dx_N
\exp \Big( - \frac{S_E^{(0)} + S_E^{(1)}}{\hbar} \Big)
\end{equation}
where $Z_{\mu}$ is the partition function of the oscillator. It is
understood that $x_0 = x_N$. Furthermore,
$S_E^{(0)}$ is given by the standard expression, while
\begin{equation}
\label{eq34}
S_E^{(1)} = \frac{\epsilon^2}{2} \sum_{i,j=1}^N
\alpha_{\mu} (\tau_i -\tau_j)
\Big[ g(x_{i-1}) - g(x_{j-1}) \Big]^2 \; .
\end{equation}
Here $\alpha_{\mu} (\tau_i -\tau_j)$ is given by an expression similar
to Eq.\ (\ref{eq9}), namely
\begin{equation}
\label{eq34a}
\alpha_{\mu} (\tau) = (2\beta)^{-1} \sum_{\omega}^{\rm{1st \; BZ}}
              {\rm e}^{-i\omega\tau}
              \frac{1}{[\omega^2]_\epsilon + \omega_\mu^2}
\end{equation}
where
$[\omega^2]_\epsilon \equiv 2[1-\cos(\omega\epsilon)]/\epsilon^2$; the
summation
is restricted to, say, the first Brillouin zone, i.\,e.\
$n = -N/2+1 \dots N/2$ (assuming $N$ to be even).

An alternative version of $S_E^{(1)}$ can be obtained directly
from the transformed Hamiltonian (\ref{eq19a}), with the
following result:
\begin{eqnarray}
\label{eq35}
{\tilde S}_E^{(1)} & = &
\frac{\epsilon}{2\omega_\mu^4} \sum_{i=1}^N g_1^2 ({\bar x}_i)
\Big(\frac{x_i - x_{i-1}}{\epsilon} \Big)^2 
       \nonumber \\
& & - \frac{\epsilon^2}{\omega_\mu^4} \sum_{i,j=1}^N 
\gamma_{\mu} (\tau_i -\tau_j)
g_1 ({\bar x}_i) g_1 ({\bar x}_j) 
\frac{x_i - x_{i-1}}{\epsilon}
\frac{x_j - x_{j-1}}{\epsilon} \; .
\end{eqnarray}
In this form, the 
coordinate-dependent mass correction, $m_1({\bar x}_i) = 
g_1^2 ({\bar x}_i) / \omega_\mu^4$, is immediately identified;
compare Eq.\ (\ref{eq20}). 
The appearance of the mid-points in this expression
is related to the approximation
$g(x_j) - g(x_{j-1}) \approx g_1 ({\bar x}_j) \cdot (x_j - x_{j-1})$
in an intermediate step.

The kernel $\gamma_{\mu} (\tau_i -\tau_j)$ can be
related to $\alpha_{\mu} (\tau_i -\tau_j)$ by noting the relations
\begin{equation}
\label{eq36}
\alpha_{\mu} (0) = \frac{1}{2\omega_\mu^2} \, ; \;
\alpha_{\mu} (\omega) - \alpha_{\mu} (0) = 
- \frac{[\omega^2]_\epsilon \alpha_{\mu} (\omega)}{\omega_\mu^2} \, ; \;
\gamma_{\mu} (\omega) = [\omega^2]_\epsilon \alpha_{\mu} (\omega) \; ,
\end{equation}
which imply  
\begin{equation}
\label{eq37}
\gamma_{\mu} (\tau_i -\tau_j) = \delta_{ij} / 2\epsilon
              - \omega_\mu^2 \alpha_{\mu} (\tau_i -\tau_j) \; . 
\end{equation}
Using (\ref{eq36}) and
(\ref{eq37}), the correspondence
between $S_E^{(1)}$ and ${\tilde S}_E^{(1)}$ can be established directly.
A more compact form also follows easily:
\begin{equation}
\label{eq38}
{\tilde S}_E^{(1)}  = 
  \omega_\mu^{-2} \sum_{i,j=1}^N \alpha_\mu (\tau_i -\tau_j) \,
  [g ({x}_i) - g ({x}_{i-1})] \, 
  [g ({x}_j) - g ({x}_{j-1})] \; .
\end{equation}
For zero temperature, $\alpha_\mu (\tau_i -\tau_j)$ 
can be found
by an elementary integration.

\subsection{Linearly coupled harmonic oscillators}
In order to illustrate some aspects of the above results, we consider
the special case of a linear coupling (i.\,e.\ $x$-independent mass
renormalisation: $g(x) = g_1 \cdot x, m_1 = g_1^2 / \omega_\mu^{4}$). 
In this case, the Fourier
representation of the action is useful:
\begin{equation}
\label{eq39}
{\tilde S}_E^{(1)} = \omega_\mu^{-2} g_1^2 \, \beta^{-1} \sum_{\omega} \;
[\omega^2]_\epsilon \, \alpha_\mu (\omega) \, | x(\omega) |^2 \; .
\end{equation}
In addition, we assume that the ``particle'' is a harmonic oscillator
of frequency $\omega_0$. Then the following (standard) result is 
easily confirmed ($\epsilon\to 0$):
\begin{equation}
\label{eq40}
Z = Z_0 \cdot Z_\mu \cdot \prod_{\omega} \bigg[
\frac{(\omega^2 + \omega_0^2)(\omega^2 + \omega_\mu^2)}
     {(\omega^2 + \omega_a^2)(\omega^2 + \omega_b^2)} \bigg]^{1/2}
\end{equation}
where $Z_0$ and $Z_\mu$ are the partition function of the oscillator
``0'' and ``$\mu$'', respectively. In addition,
$\omega_a$ and $\omega_b$ denote the eigen-frequencies of the 
coupled two-oscillator system, which obey the following relations:
\begin{equation}
\label{eq41}
\omega_a^2 \cdot \omega_b^2 = \omega_0^2 \cdot \omega_\mu^2 \; ; \;\;
\omega_a^2 + \omega_b^2 = \omega_0^2 + \omega_\mu^2 
           + (m_1 / m)  \omega_\mu^2 \; .
\end{equation}
In particular, considering $m_1 / m$ fixed and $\omega_\mu\to\infty$,
we find $\omega_a$ and $\omega_b$ to be given by
$\omega_0 (1+m_1/m)^{-1/2}$ and 
$\omega_\mu (1+m_1/m)^{1/2}$, respectively. Naturally, as long as
$\omega_0 \ll \omega_\mu$, the low-temperature ($k_B T\ll\hbar\omega_\mu$)
thermodynamics is determined by the low-frequency oscillator, implying
an enhancement of the specific heat due to the coupling.
Accordingly, the low-frequency
response,\footnote{A detailed 
  (perturbative) study of the mobility of
  ``polaronic objects'' is given in Appendix C of \cite{eckern89},
  including the ``large'' polaron. As can be seen from the
  polaron case, the self-energy-mass and transport-mass enhancement is
  generally not the same.}
which is easily deduced from the above 
results, contains
the inverse of $[(-i\omega +0)^2 (1+m_1/m) + \omega_0^2]$, and hence 
also displays the mass enhancement $m \to m(1+m_1/m)$.

\subsection{Perturbation theory: quantum mechanics}
A direct approach to the questions under discussion is given by
quantum-mechanical perturbation theory, provided the coupling between
particle and oscillator is small; this approach was also used in
\cite{shimshoni89} for weakly coupled superconductors. 
We write ${\tilde H} = H_0 + H_1$,
see Eq.\ (\ref{eq19a}), where $H_0$ is the Hamiltonian of the particle
plus the oscillator; the coupling is given by
\begin{equation}
\label{eq51}
H_1 = \frac{1}{2m\omega_\mu^2} \Big( {\hat p} g_1 ({\hat x}) +
      g_1 ({\hat x}) {\hat p} \Big) {\hat p}_\mu
      + \frac{1}{2m\omega_\mu^4} [g_1 ({\hat x})]^2 {\hat p}_\mu^2 \; .
\end{equation}
The eigen-energies and -functions of $H_0$ are assumed to be given,
\begin{equation}
\label{eq52}
E_{n,{\ell}}^{(0)} = \varepsilon_n + \hbar\omega_\mu (\ell + 1/2) \; ,
\end{equation}
and we consider the $\{ \varepsilon_n \}$ to be non-degenerate. Furthermore,
we will assume that the level spacing of the particle
is small compared to $\hbar\omega_\mu$. The eigen-functions are product states, 
$|n,\ell\rangle = |n\rangle \otimes |\ell\rangle$, where $|\ell\rangle$
refers to the harmonic oscillator functions.
Using standard perturbation theory, we find in first order
\begin{equation}
\label{eq53}
\Delta E_{n,{\ell}}^{(1)} = 
     \hbar\omega_\mu (\ell + 1/2) \cdot \langle n| m_1({\hat x}) |n\rangle
     / 2m \; , \;\; m_1({\hat x}) = [g_1 ({\hat x})]^2 /\omega_\mu^4 \; ,
\end{equation}
which corresponds to an $x$-dependent frequency enhancement of the oscillator
(as discussed in the previous subsection for a constant frequency
enhancement). In second order, we encounter the energy denominator
$\hbar\omega_\mu\Delta\ell + \varepsilon_n - \varepsilon_k$
with (i) $\Delta\ell = 0$, (ii) $\Delta\ell = \pm 1$, and 
(iii) $\Delta\ell = \pm 2$. The contribution (i) corresponds to the
frequency enhancement mentioned above in second order perturbation 
theory in the particle
subspace, and is of no interest here. Considering (ii) and (iii), we
concentrate on the low-lying
energies so that $\varepsilon_n\ll\hbar\omega_\mu$,
$|\varepsilon_n - \varepsilon_k|\ll\hbar\omega_\mu$. Hence we neglect
$\varepsilon_n - \varepsilon_k$ in the denominator and assume that
the {\em restricted} sum $\sum_k| k\rangle\langle k|$ is a good
approximation to the unity operator in the particle subspace;
this assumes that there is a sufficient
number of particle-energies in the relevant energy range. The contribution
(iii) is also related to the $x$-dependent frequency correction of
the oscillator, in second order and hence $\sim m_1^2$, and will be
neglected as well. Finally, the dominant contribution (ii), $\sim m_1$, 
is given by
\begin{equation}
\label{eq54}
\Delta E_{n,{\ell}}^{(2),{\rm (ii)}} 
    =  - \frac{1}{8m^2\omega_\mu^4}
   \langle n| \Big( {\hat p} g_1 ({\hat x}) 
               + g_1 ({\hat x}) {\hat p} \Big)^2 |n\rangle \; .
\end{equation}
Clearly, Eqs.\ (\ref{eq53}) and (\ref{eq54})
are consistent with the results discussed in the previous
subsection. We conclude that the effective correction to the
particle Hamiltonian, in this order, is given by
\begin{eqnarray}
\label{eq54a}
H_1^{\rm eff} & = & \frac{\hbar\omega_\mu}{4m} m_1({\hat x})
                - \frac{1}{8m^2\omega_\mu^4}
       \Big( {\hat p} g_1 ({\hat x})+ g_1 ({\hat x}) {\hat p} \Big)^2
       \nonumber \\
       & = & \frac{\hbar\omega_\mu}{4m} m_1({\hat x})
             - \frac{m_1}{2m^2} {\hat p}^2   
             + \frac{i\hbar m_1^\prime}{2m^2} {\hat p}
             + \frac{\hbar^2 m_1^{\prime\prime}}{8m^2}
	     - \frac{\hbar^2 (g_1^\prime)^2}{8m^2\omega_\mu^4} \; .
\end{eqnarray}       
On the other hand, consider the pct- and Weyl-Hamiltonian,
(\ref{eq24}) and (\ref{eq29}), 
with $m({\hat x}) = m + m_1({\hat x})$, in the limit
of small $m_1$. In linear order in $m_1$, $H_{\rm pct}$ and
$H_{\rm W}$ agree, with the result
\begin{equation}
\label{eq55}
H_{\rm pct,W} \simeq H_{0} - \frac{m_1}{2m^2} {\hat p}^2
              + \frac{i\hbar m_1^\prime}{2m^2} {\hat p}
	      + \frac{\hbar^2 m_1^{\prime\prime}}{8m^2} \; ;
\end{equation}
here $H_{0} ={\hat p}^2/2m + V({\hat x})$. Obviously,
Eqs.\ (\ref{eq54a}) and (\ref{eq55}) are not consistent; only the terms
proportional to ${\hat p}^2$ and ${\hat p}$ agree. Most notable is
the large contribution due to the frequency renormalisation of the
oscillator, compare (\ref{eq53}), which is larger than the term
(\ref{eq54}) by a factor of the order of 
$\hbar\omega_\mu / \langle n| {\hat p}^2/m |n \rangle$.

\subsection{Perturbation theory: path integral}
Within the path integral formalism, the first-order correction to
the free energy is given by
\begin{equation}
\label{eq61}
\Delta F^{(1)} = \beta^{-1} \langle {\tilde S}_E^{(1)} \rangle_0
\end{equation}
where the average is with respect to the particle action alone. In order
to evaluate this expression, we need the correlation function
\begin{equation}
\label{eq62}
  \chi_{gg} (\tau - \tau^\prime) \equiv
  \langle g (x(\tau)) g (x(\tau^\prime) \rangle_0
  \;\; \to \;\;
  \langle {\cal T}_\tau 
  \Big[ g ({\hat x}(\tau)) g ({\hat x}(\tau^\prime) \Big] \rangle_0 \; ,
\end{equation}
where ${\cal T}_\tau$ is the time ordering operator. We express
this correlation function with the help of the 
eigen-energies and -states, $\varepsilon_n$ and $|n\rangle$, introduced
above, with the result
\begin{equation}
\label{eq63}
  \chi_{gg} (\tau - \tau^\prime) = Z_0^{-1} 
  \sum_{n,k} {\rm e}^{-\beta\varepsilon_n/\hbar} \,
  | \langle n | g({\hat x}) | k \rangle |^2 \,
  {\rm e}^{| \tau - \tau^\prime | (\varepsilon_n - \varepsilon_k)/\hbar}
  \; 
\end{equation}
where $Z_0 = \sum_n \exp (-\beta\varepsilon_n/\hbar)$.
Thus we find in the limit $\epsilon\to 0$ 
(compare Eq.\ (\ref{eq38})):
\begin{eqnarray}
\label{eq64}
\Delta F^{(1)} & = & \beta^{-1} \omega_\mu^{-2}
  \int_0^\beta \! d\tau \int_0^\beta \! d\tau^\prime \;
  \alpha_\mu (\tau -\tau^\prime ) \;
  \frac{\partial^2}{\partial\tau\partial\tau^\prime}
  \chi_{gg} (\tau - \tau^\prime)
  \nonumber \\
  & = & \beta^{-1} \omega_\mu^{-2} \sum_\omega
  \omega^2 \, \alpha_\mu (\omega) \, \chi_{gg} (\omega) \; .
\end{eqnarray}
The frequency sum is evaluated using standard techniques; we consider
the limit $\beta\omega_\mu \gg 1$, and assume as above 
that the relevant energy differences 
$(\varepsilon_n - \varepsilon_k)^2$ are small compared to
$(\hbar\omega_\mu)^2$. The result has the expected form,
\begin{equation}
\label{eq65}
\Delta F^{(1)} = Z_0^{-1} \sum_n {\rm e}^{-\beta\varepsilon_n /\hbar}
\Delta E_n  \; ,
\end{equation}
where
$\Delta E_n = \langle n| H_1^{\rm eff} |n \rangle$
in agreement with Eqs.\ (\ref{eq53}) and (\ref{eq54}); see also
(\ref{eq54a}).

To summarise briefly the results of Sects.\ 4.5 and 4.6,
perturbation theory thus shows that there is an effective
potential contribution $\sim m_1$  which can be interpreted as an
$x$-dependent modulation of the ground state energy of the oscillator.
Furthermore, the second term in $H_1^{\rm eff}$ is {\em not} consistent with 
what is predicted from the pct- or Weyl-Hamiltonian.

\subsection{Variational approach}
For several decades projector methods have been known to be a
useful tool for deriving an effective description for a complex
quantum system; recall e.\,g.\ Feshbach's approach \cite{feshbach58} 
to nuclear reactions,
and the Mori-Zwanzig formalism \cite{zwanzig61} in statistical physics.
For the present case, we wish to derive an
effective Hamiltonian defined as projection onto the ground state of
the oscillator, eliminating its excited states. In close analogy to
Feshbach's method, we use the following ansatz for the wave-function
of the system, i.\,e.\ particle plus oscillator:
\begin{equation}
\label{eq71}
|\Psi\rangle = \sum_{\ell = 0}^L \, |\psi_\ell\rangle\otimes |\ell\rangle \; .
\end{equation}
Employing the variational principle of quantum mechanics, we obtain
\begin{equation}
\label{eq72}
\sum_{\ell^\prime = 0}^L ({H})_{\ell,\ell^\prime} \,
       |\psi_{\ell^\prime}\rangle = E \, |\psi_\ell\rangle 
\end{equation}
and, after eliminating $|\psi_\ell\rangle$ with $\ell = 1,2,\dots,L$ from
this set of equations, an effective $E$-dependent Hamiltonian can be defined
by
\begin{equation}
\label{eq73}
H_{\rm eff}^{\rm (L)} (E) \, |\psi_0\rangle = E \, |\psi_0\rangle \; .
\end{equation}
Applying this procedure to the Hamiltonian given in Eq.\ (\ref{eq19a}),
and taking $L=1$ for simplicity, we find
\begin{eqnarray}
\label{eq74}
H_{\rm eff}^{(1)} (E) & = & H_{00} + 
     \frac{1}{2} \hbar\omega_\mu \Big( 1 + \frac{m_1 ({\hat x})}{2m} \Big)
     \nonumber \\
& &     - \frac{\hbar}{8m^2\omega_\mu^3} 
     \Big( {\hat p}g_1({\hat x}) + g_1({\hat x}){\hat p} \Big)
     X_3^{-1} \Big( {\hat p}g_1({\hat x}) + g_1({\hat x}){\hat p} \Big)
\end{eqnarray}
where $H_{00} = {\hat p}^2/2m + V({\hat x})$, and
\begin{equation}
\label{eq75}
X_3 = H_{00} + \frac{3}{2}\hbar\omega_\mu 
             \Big( 1 + \frac{m_1 ({\hat x})}{2m} \Big) - E \; .
\end{equation}
From these expressions, the perturbative results of the previous
subsection follow easily, using the approximation $E\simeq\hbar\omega_\mu/2$,
hence $X_3 \simeq \hbar\omega_\mu$, which is valid under the same conditions
as discussed above. At present, it is unclear whether Eq.\ (\ref{eq74})
would give meaningful results beyond this limit. Furthermore, a preliminary
study of $H_{\rm eff}^{(2)} (E)$ indicates that it is not straightforward 
to derive concise
conclusions within this method, for the most relevant case
$m_1 \sim m$.

\section{Conclusion}
\label{con}
In this article we have studied several aspects of effective quantum 
theories, arising when a system is coupled to an environment and the 
latter is integrated out. We considered, in particular, the situation
where the coupling between system and environment is non-linear: for
the model case of a particle coupled to a single\footnote{All results
are easily generalised to the case where the coupling is to several
oscillators.} high-frequency oscillator, this non-linear coupling
effectively leads to a coordinate-dependent mass of the particle, on
the classical level and for low frequencies.

However, on the quantum level, we have not been able 
to confirm the assertion that an effective low-energy
theory can be formulated in which the coupling to the
environment and the properties of the environment are expressed through
the coordinate-dependent mass alone. This becomes apparent when
considering the (discrete) path integral formulation, since high-frequency
contributions to the effective action cannot simply be discarded.
Consistently, it is easily seen in perturbation theory that due to the
non-linearity, significant and model-dependent contributions arise
which reflect the back-action of the particle on the environment.

\begin{acknowledgement}
  UE particularly acknowledges the
  close and fruitful cooperation, as well as many discussions
  on some aspects of this work, with Vinay Ambegaokar and Gerd Sch\"on.
  We are grateful to Hajo Leschke for comments, and to Hagen Kleinert for
  pointing out relevant references.
\end{acknowledgement}

\end{document}